\documentclass[manuscript=true, review=false, anonymous=false, nonacm=true]{acmart}

\usepackage{todonotes}
\usepackage{enumitem}
\usepackage{microtype}
\usepackage{graphicx}
\usepackage{subfigure}
\usepackage{booktabs} 
\usepackage{hyperref}
\usepackage{latexsym}
\RequirePackage{fix-cm}
\usepackage{mathptmx}      
\usepackage{relsize}
\usepackage{footnote}
\makesavenoteenv{tabular}
\makesavenoteenv{table}
\usepackage{threeparttable}
\usepackage{amsmath, mathtools}
\usepackage{array, url}
\usepackage{dblfloatfix}
\usepackage{tcolorbox}
\usepackage{balance}
\usepackage[framemethod=TikZ]{mdframed}
\graphicspath{{figures/}}
\DeclareGraphicsExtensions{.pdf,.jpg,.png}

\mdfsetup{
    roundcorner=10pt,
    innerleftmargin=20pt,
    innerrightmargin=20pt,
    innertopmargin=5pt,
    innerbottommargin=10pt,
    frametitleaboveskip=10pt,
    font=\small
}

\mdfdefinestyle{greybox}{
    skipabove=10pt,
    skipbelow=20pt,
    middlelinecolor=gray!50,
    middlelinewidth=1.5pt,
    backgroundcolor=gray!5
}

\mdfdefinestyle{pinkbox}{
    middlelinecolor=purple!75!black,
    middlelinewidth=1.5pt,
    backgroundcolor=purple!5!white
}

\usepackage{xpatch}

\makeatletter
\xpatchcmd{\ps@firstpagestyle}{Manuscript submitted to ACM}{}{\typeout{First patch succeeded}}{\typeout{first patch failed}}
\xpatchcmd{\ps@standardpagestyle}{Manuscript submitted to ACM}{}{\typeout{Second patch succeeded}}{\typeout{Second patch failed}}    \@ACM@manuscriptfalse
\makeatother

\AtBeginDocument{%
  \providecommand\BibTeX{{%
    \normalfont B\kern-0.5em{\scshape i\kern-0.25em b}\kern-0.8em\TeX}}}

\begin{document}

\title[Design Considerations for Data Daemons]{Design Considerations for Data Daemons: Co-creating Design Futures to Explore Ethical Personal Data Management}

\author{Wiebke Toussaint}
\authornote{Corresponding author}
\email{w.toussaint@tudelft.nl}
\author{Olya Kudina}
\author{Aaron Yi Ding}
\affiliation{%
  \institution{Technology, Policy \& Management, TU Delft}
  \country{The Netherlands}
}

\author{Alejandra Gomez Ortega}
\author{Jered Vroon}
\author{Jacky Bourgeois}
\affiliation{%
  \institution{Industrial Design Engineering, TU Delft}
  \country{The Netherlands}
}

\author{Julian Harty}
\orcid{0000-0003-4052-0054}
\affiliation{%
  \institution{Commercetest Ltd., The Open University}
  \country{United Kingdom}
}

\author{G\"urkan Solmaz}
\orcid{}
\affiliation{%
  \institution{NEC Laboratories Europe}
  \country{Germany}
}

\author{Ella Peltonen}
\orcid{}
\affiliation{%
  \institution{University of Oulu}
  \country{Finland}
}

\renewcommand{\shortauthors}{Toussaint et al.}

\begin{abstract}
Mobile applications and online service providers track our virtual and physical behaviour more actively and with a broader scope than ever before. This has given rise to growing concerns about ethical personal data management. Even though regulation and awareness around data ethics are increasing, end-users are seldom engaged when defining and designing what a future with ethical personal data management should look like. We explore a participatory process that uses design futures, the Future workshop method and design fictions to envision ethical personal data management with end-users and designers. To engage participants effectively, we needed to bridge their differential expertise and make the abstract concepts of data and ethics tangible. By concretely presenting personal data management and control as fictitious entities called Data Daemons, we created a shared understanding of these abstract concepts, and empowered non-expert end-users and designers to become actively engaged in the design process. 

\end{abstract}



\maketitle

\section{Introduction}
\label{introduction}

Connected digital devices, including laptops, smartphones, wearables, and diverse Internet of Things (IoT) devices, produce ongoing personal tracking data streams that track what, when, where and with whom we move, interact, and socialise. Personal tracking data have become important drivers of mobile and digital services, with businesses even viewing them as the "new gold"~\cite{moon2019cultivated}. Consequently, our virtual and physical behaviour is tracked more actively and with a broader scope today than ever before. This is seldom in the interest of individual end-users, even if they voluntarily accept the risks~\cite{phelan2016creepy}. Ethical concerns pertaining to the processing and management of personal data have thus moved to the forefront of policy debate. However, translating ethical guidelines and policies into regulatory mechanisms and standards remains a challenge~\cite{mittelstadt2016algorithms}. Similarly, ethical concerns do not translate readily into design considerations for data-driven products and services. Despite increasing awareness around data ethics, and regulation around the use and management of personal data, many people have resigned themselves to the belief that their personal data is more likely to be abused than to serve them~\cite{phelan2016creepy, colnago2020informing}. Even though end-users are most likely to be adversely impacted by extractive data practices, participatory approaches to ethical personal data management remain scarce~\cite{whitman2018potential}. The abstract nature of data~\cite{vandenberghe2016anthropomorphism} and differential expertise of participants~\cite{baumer2017toward} are considered key barriers to engaging end-users in the design process of data-driven applications. High data and digital literacy requirements often exclude both application designers and end-users from actively participating in the design of ethical data-driven services~\cite{roeck2011infusing}. 

This paper presents a first exploration of how the perspectives of end-users and designers can inform the future development of ethical personal data management. Using design futures and a participatory approach, we overcome the barriers presented by abstraction and the inherent data and digital literacy requirements of data ethics with a fictitious concept that we call Data Daemons. The Data Daemons enabled us to co-create design fictions and design considerations that make ethical personal data management more concrete and reflective of the values of a group of expert and non-expert participants. We see this as a valuable contribution that builds on participatory design from the HCI community to bring end-user and designer perspectives into ethical personal data management. 

\section{Background and Related Work}

\subsection{The Status Quo of Personal Data Management}

Collecting and processing personal tracking data can be highly invasive and seldom in the best interest of end-users~\cite{Goodman2014}. Many online services and mobile applications harvest and trade personal tracking data with a multitude of data brokers and third-party service providers. As Helles et al.~\cite{helles2020infrastructures} show, the online tracking business feeds an attention economy in which end-users are more unsuspecting subjects than collaborating partners. 

Treating end-users as subjects is contested in the data science domain. Metcalf and Crawford~\cite{metcalf2016where} point out the growing disconnect between the limited ethical considerations in data science research on the one hand, and established tools and practices of research ethics in disciplines that traditionally conduct human subject research on the other hand. Ethical challenges arise, in part, from data science methods making data "infinitely connectable, indefinitely repurposeable, continuously updatable and ... removed from ... context"~\cite{metcalf2016where}. These same conditions inherently dominate the nature of personal tracking data. Risks arising from unethical data practices are not limited to traditional physical harms, but include concerns like the rise of new digital divides, infringement on information privacy, wrong conclusions due to decontextualisation, and misleading claims of objectivity and accuracy~\cite{boyd2012critical}.   

\subsection{Participatory Design Perspectives on Personal Data and Algorithms}

Participatory Design is an established tradition in the design of information and communication technologies~\cite{robertson2013particpatory}. It stipulates that the users of technology must be able to influence its design, even if they have no expertise or prior knowledge in technology design. Furthermore, it encourages mutual learning between users and designers, acknowledging that users who lack expertise in technology design may not know what they want without knowing what is possible. Participatory Design has been used to engage end-users on personal data. For example, Rosenbank et al.~\cite{rosenbank2016design} present an art-based approach that involves participants in co-designing their digital shadows with extracts of personal metadata. In Metadating~\cite{elsden2016metadating}, Elsden et al. take a speculative approach to explore how people use personal data to communicate their identity in a future-focused speed-dating event. Looking at design interventions, Vandenberghe and Slegers~\cite{vandenberghe2016anthropomorphism} use anthropomorphism as a strategy to learn about the personal health data information needs of end-users during the ideation phase of a future health data application. Colnago et al.~\cite{colnago2020informing} conduct semi-structured interviews to elicit the opinions of potential users on different privacy-related design configurations to inform design considerations for personalized privacy assistants. Baumer~\cite{baumer2017toward} highlights the need for Participatory Design in algorithm design, but also points to the challenge that engaging with algorithm design requires significant expertise, which participants typically do not have. 

While promising, participatory research on ethical personal data management and algorithms is still an emerging field~\cite{whitman2018potential}. In particular, Whitman et al. note that most participatory research, like the research discussed above, focuses on facilitating participants' engagement with data, or centering end-users in the process of developing effective and engaging tools. Participatory research that focuses on values and ethics of personal data management and algorithm design is scarce. 

\subsection{Using Fictions and Futures to Explore What Could and Should Be}

Design futures are a strategy for actively recruiting the future into design practice~\cite{reeves2016future}. Reeves et al. point out that to be of value, design futures need to be considered legitimate by a community that extends beyond the visionaries. They suggest that participatory approaches and in particular design fictions can be used to enhance the legitimacy of design futures by enabling broader participation in envisioning the future. The Future workshop~\cite{vidal2005future} is a facilitated approach that guides participants through a process of co-creating fantasy futures. It was developed as a form of structured citizen engagement, and emphasises group dynamics, collective work and collaborative learning, and facilitated, creative problem solving. Design fictions invite the narrative and future-oriented characteristics of science fiction into the practical and material world of design~\cite{bleecker_2009_design}. They are widely used in HCI to explore future scenarios, to map out design possibilities, to engage users in participatory design, or to offer critique through story-telling and character development~\cite{baumer_2020_evaluating}. We initiated our participatory research with the Future workshop to invite participants into a co-creative process. Design fictions emerged as a participatory tool later on to continue engaging participants in the concepts that were inspired by the workshop. 
\section{Co-creating Data Daemons}
\label{approach}

We approached this study as an exploratory and emergent process, with a focus on active participation and co-creation. Our starting point was the goal to explore design futures for ethical personal data management in an online Future workshop. The Future workshop brought together a group of 11 participants to scope the problem space, envision ideal fantasy futures and identify barriers towards them. The participants were a demographically heterogeneous group of researchers with 10 different nationalities from 4 continents, 6 male and 5 female researchers, and an age span of two decades. Participants' professional backgrounds were in both technical and non-technical disciplines. While all participants had an interest in ethical personal data management, expertise was clustered around specific aspects of the topic -- ethics, personal data, data management and data-driven products and services. Some participants self-identified as non-experts, bringing perspectives from design, industry, and their personal lives. The output of the workshop can be viewed online\footnote{\url{https://miro.com/app/board/o9J_kkfAKjE=/}}.

\subsection{Emergence of Data Daemons}
During the workshop two challenges became evident: firstly, because future visions emerge from experienced reality, ideas were limited by existing data management practices, technologies and paradigms, which at best incorporate ethics as an afterthought. While we expected alternative futures to exist, we needed a way to inspire a “what if” perspective to provoke them. Secondly, we realised that building fantasy futures of abstract concepts like ethics and data was difficult, as participants had different backgrounds and a vastly different understanding of what these concepts mean. To overcome these differences we needed a bridge that could connect different standpoints. 

These challenges inspired us to conceptualise Data Daemons. Data Daemons would be owned by individuals, and have the responsibility of managing and controlling their owner’s personal data in their owner’s best interest. They merge the functionality of daemons found in computer programmes, and daemons invoked in the popular fantasy series "His Dark Materials" by Sir Philip Pullman. In computer science, daemons are programmes that run routine background processes and service requests in multitasking operating systems. They operate independently without user intervention, and can interact with other processes. Pullman's fantasy daemons are animal-shaped, physical manifestations of our inner self. They are spiritual guides that are at the same time external to their humans, yet a part of them. Combining these two kinds of daemons led us to Data Daemons: artificial entities that know us intimately and perform data management tasks on our behalf. Data Daemons manage all aspects of our personal tracking data and provide useful services like the daemons that already execute tasks in computing systems. In addition, they also concretise relationships between people and their personal tracking data to be like the relationships between Pullman's human characters and their daemons. In our ideal future, Data Daemons thus embody the relationship between humans and the systems that collect and process our personal tracking data. They are the virtual manifestation of our digital self.

\subsection{Co-creative Approach}
This shared conceptualisation of fictitious Data Daemons created a common language for workshop participants, while also liberating them to be more ambitious and creative in imagining their ideal ethical personal data futures. The Future workshop inspired a series of follow-up meetings and writing exercises to further develop the concept of Data Daemons to explore design considerations for ethical personal data management. In the follow up meetings we selected design fictions as a tool to co-create narrative prototypes of Data Daemons that highlight design considerations for them. The design fictions were refined through several iterations. 

\begin{figure*}
    \centering
    \includegraphics[width=0.8\linewidth]{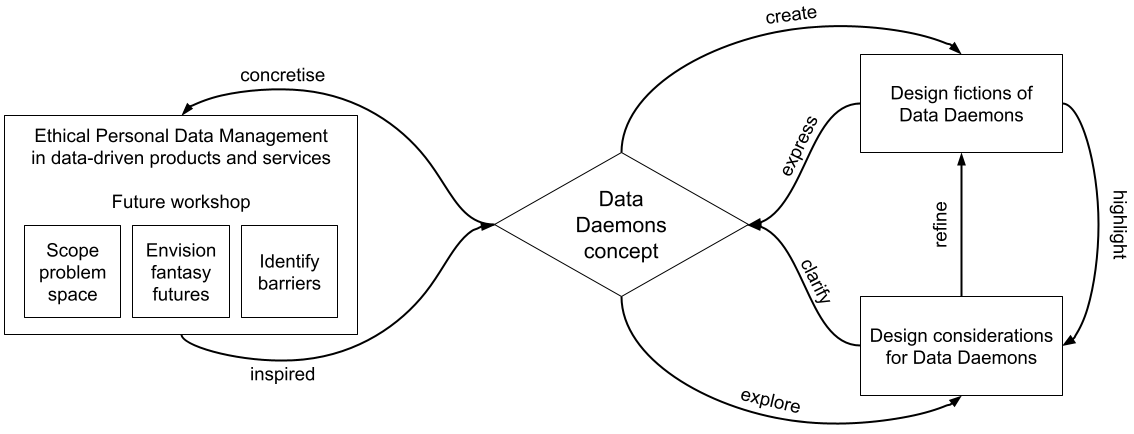}
    \caption{Co-creative approach to developing Data Daemons}
    \label{fig:datadaemonsapproach}
\end{figure*}

The co-creative process that emerged is shown in Figure \ref{fig:datadaemonsapproach}, which depicts the interdependent relationship between the Data Daemon concept, design fictions of Data Daemons, and design considerations for Data Daemons. We used the Data Daemons concept to explore design considerations and create design fictions. The design fictions highlight design considerations. Reflecting on the design considerations helped to refine the design fictions over several iterations. With each iteration the design fictions and design considerations became more nuanced, expressing and clarifying the Data Daemon concept with greater detail, and with it concretising participants’ expectations for ethical personal data management of data-driven products and services. Next we present the design fictions as refined in the final iteration.

\section{Design Fictions of Data Daemons}
\label{design_fictions}

The design fiction scenarios presented here are fictional reflections on personal Data Daemons by three personae with different values and concerns that are influenced by their data and digital literacy. Proactive Pippa (Section \ref{sec:fictionPowerUser}) is highly data-conscious and an Internet native. Mother Maria (Section \ref{sec:fictionNonDigitallyNativeUser}) feels overwhelmed by the rapid pace of digital change. Middle-aged Regular Robert (Section \ref{sec:fictionRegularUser}) captures the middle ground between the two. The design fictions demonstrate how people with varying degrees of data literacy reflect on their interaction with, and experience of, their Data Daemon. 

\subsection{Design Fiction 1: Pippa improves her performance and health with her data}
\label{sec:fictionPowerUser}
\small{\texttt{PIPPA: Data-conscious Internet native in her late twenties. Wary of fake news and social media manipulation. Suffers from stress and migraines. Into mindfulness.}}

\smallskip
\textit{"Pivo, my data concierge, knows me intimately. He knows what I want, and guards my preferences. I request them from him on a monthly basis to help me reflect on my behaviour and activities. Pivo can summarise whether I have exercised enough, who I have seen the most, and where I have been. Based on that, he recommends what I should change in the coming month to meet my personal and career goals. Of course I don't have to accept his recommendations, and I sometimes reject them to ensure that I can still make independent decisions. Since working with Pivo I have finally managed to prioritise self-care. He infers my default data collection and processing options for all my digital services based on my monthly reflections. Pivo highlights data donation opportunities that fit my goals, for example, donating data to inform the design of a new product that can help me cope with migraines. Recently, the loud city noises and weather conditions have intensified my migraines. When I got on a busy bus last Tuesday, I was automatically guided to a quiet seat away from the harsh sunlight. Thanks to Pivo, the data from my various health monitors are now combined with hyper-local weather forecasts and the public transport systems to plan my route, minimising adverse effects. Pivo has arranged that just enough data is revealed on a temporary basis with the weather app and transit systems. The shared data automatically expires at the end of my journey. I get the benefit of personalised route planning, while the various systems can calibrate and improve their accuracy based on real life conditions. My health insurance actually contributes to the cost of Pivo and the services he negotiates, because this is much more effective \emph{and} significantly cheaper than medical care. No-one else needs to access my sensitive data, least of all my employer! Work pays the balance of Pivo’s subscription because I'm able to do better work as a result of the services he facilitates. Keeping me healthy is still good business value."}

\subsection{Design Fiction 2: Robert's data empowers him to do more and do good}
\label{sec:fictionRegularUser}

\small{\texttt{ROBERT: Mid-career professional in the EU. Discusses gadgets and tech trends with colleagues. Reads Hacker News. Expects tech to serve him and likes to feel good.}}

\smallskip
\textit{"In the early days, when Ruba, my Data Daemon, was not yet fully trained, it would ask for my preferences almost daily. But now, I barely notice Ruba anymore. Sometimes it checks in briefly to confirm we are still aligned, which we usually are. Ruba's services extend well beyond managing my personal data ownership of course, as we leave traces all over the place. We really established our relationship when I got my smart washing machine. Ruba guided me through the setup and interpreted the options for sharing my data. It took less time than setting up my sharing preferences on social media. When cookies started to be introduced on my home devices, I rejected all of them at first, just like I did with Internet browsers. I then realised that by accepting certain cookies, Ruba was able to compile reviews and perspectives that contrast my own to broaden my horizon. Ruba quickly adapted to accept useful cookies on my behalf. And when cookies collect data for targeted advertising, Ruba skillfully haggles good returns and ensures that the advertisements I am shown do not make me feel like I've sold my soul. I even donate some of my ad-view revenue towards planting trees. Ruba suggested that I share my anonymised speech data, even for commercial purposes, because it knows that I like openness and believe in win-win situations. Donating data helps improve speech recognition for everyone -- and I get a small compensation that I usually use for a cup of coffee. Thanks to Ruba's excellent negotiations I strike a fairly even balance, paying only a little for some digital services each month. For example, I pay to get unrestricted access to a selection of curated websites, including two news platforms, without any advertisements. However, I give free access to my anonymised home energy usage data to contribute towards a more sustainable world."} 

\subsection{Design Fiction 3: Maria benefits from her smart services with peace of mind}
\label{sec:fictionNonDigitallyNativeUser}
\small{\texttt{MARIA: Mother of two teenage daughters. Overwhelmed by the rapid pace of digital change. Seeks stability and connection to the familiar. Chicken curry champion (family secret).}}

\smallskip
\textit{"I have to confess, I did not want a Data Daemon. I was scared by the idea of having to interact with something so vaporous. But being digitally disconnected really isolated me from my family. So my daughter convinced me to try Myra. Myra is a tangible Data Daemon, with a minimalist touchscreen interface that allows me to effectively manage my interactions with her. Myra does not bother me for tiny tasks, only important events and emergencies. She is my anchor in the digital world, which is very foreign to me. Myra ensures that things run smoothly, always. First, Myra helped me go through the privacy settings on my smartwatch, the only smart device I own. I usually get anxious and confused with PC software updates asking to share my data, so I had a lot of questions about terminology and functionality. Myra gently guided me, gave insightful tips and responded to all my questions, patiently. She even helps me to digest complicated, long, consent forms! Nowadays I trust Myra and feel relieved I no longer have to navigate the digital world alone. Myra keeps track of the data from my smart watch and my daughter says I can double-check her decisions. I could even ask Myra to step aside if I wanted to be in control, but I would need to ask for help to learn how to do that. Myra scans the data streams for suspicious activity and catches unusual behaviour. My daughter explained that Myra takes preventative actions to protect my health tracking data, rather than alarming me unnecessarily. Myra has a good relationship with all the family's Data Daemons. She regularly fetches family updates, which makes me happy. Myra also collects news from neighbours and insights from experts, which helps me stay in touch and stay informed."}

\section{Design Considerations for Data Daemons}
\label{design_considerations}

Robert, Pippa, and Maria have diverse values, specific needs and unique ways of taking advantage of their personal tracking data. Nonetheless the design fictions have commonalities that highlight design considerations for ethical personal data management across the personae. Data Daemons clearly hold a position of power. They connect end users to opportunities that are useful for them (e.g. helping Pippa handle her migraines better) and, where possible, benefit society at large (e.g. motivating Robert to share his speech data). Across the design fictions, the Data Daemons are deeply personal, accountable, and adaptive. They interpret, discover, access, negotiate and trade personal tracking data, and facilitate interaction and collaboration. These capabilities emerged from reflections on the design considerations entailed in the design fictions. By finding a common framing for expressing our collective, desired design considerations for Data Daemons and thus clarifying our shared understanding of the Data Daemons concept, we agreed that ethical personal data management should include, as a minimum, aspects of:
\begin{enumerate}
    \item Supported personal choice and control
    \item Interaction and collaboration through negotiation
    \item Accountability and adaptation to value change
\end{enumerate}
Below we discuss how the design fictions highlight these considerations, and what this implies for ethical personal data management.

\subsection{Supported personal choice and control}
For Pippa and Maria, it is important that their Data Daemons present them with choice and control. Pippa can choose to ignore recommendations and has control over how often her Data Daemon presents feedback. She wants to avoid overreliance and retain a degree of independence. Maria chooses an embodied Data Daemon that makes interaction easier for her. As is apparent from the design fictions, exerting choice and control over personal tracking data can require considerable effort, background knowledge, insights into data usage, and familiarity with your preferences and values. The lower a person's data literacy, and the more data is generated, the more burdensome its management becomes. Consequently, choice and control cannot be realised without offering support. The Data Daemons thus automate repetitive, dynamic, and complex data management tasks and perform activities humans cannot do: encoding and tracing data usage permissions and purposes, managing temporary data sharing and expiration, and detecting, validating, and mitigating digital threats.

\subsection{Interaction and collaboration through negotiation}
The Data Daemons are integrated into their personae's existing worlds, and rely on interaction and collaboration through negotiation to help them achieve advantage in it. In the design fictions, personal tracking data has no intrinsic economic value. Rather, its value arises out of the affordances it enables for the humans that generate it, and the benefits it creates for society. This means that data is valuable in so far as it supports the development of relations between humans and their environment. Ruba offers Robert new perspectives and negotiates compensation for the usage of his data. Pivo negotiates privacy sensitive usage agreements for Pippa's data streams and integrates with the health insurance system. Myra maintains relationships with the Data Daemons of Maria's family and enables her to stay connected to the people she cares about. Discourse is a precondition for negotiation. Our design fictions propose three types of discourse for the Data Daemons: discourse with their persona, discourse with other Data Daemons, and discourse with external systems. Discourse between a persona and their Data Daemon assumes that data usage concepts can be interpreted and aligned with personal values and preferences. Discourse with external systems and between Data Daemons points to the need for a common language that can express, capture and interpret relevant data usage concepts and concerns, as well as the purposes for which data is being used. 

\subsection{Accountability and adaptation to value change}
The Data Daemons are accountable to their personae. They accompany their personae in their digital worlds, interpreting what their data means and represents in an individualised manner that is digestible and relatable. Their interactions are transparent, they offer explanations and adapt with the individual needs of their personae: Ruba asks, checks in and confirms, Pivo requests, summarises and recommends, and Myra guides gently and responds patiently. In turn, the personae have gotten accustomed to their Data Daemons offering services tailored to their unique values, needs and preferences. They trust their Data Daemons and are willing to reveal intimate information; like Pippa who has authorised Pivo to help her uncover blindspots through personalised reflection and goal-setting. Additionally, Ruba exhibits the ability to adapt to Robert over time, while Ruba and Maya offer decision-making and prioritisation support to help Robert and Maria navigate the digital world. By checking in, the Data Daemons acknowledge they may get out of sync with their persona's needs and values. This highlights that personal data management is an ongoing conversation that needs to adapt over time, rather than a once-off decision.

\section{Reflection on Co-creating Data Daemons}

We set out to explore ethical personal data management for data-driven products and services from the perspective of users and designers. Here we reflect on the overall process and the specific contributions of the Future workshop, Data Daemons, and design fictions. The main value of the work lies in exploring a process for engaging a group of heterogeneous participants in the abstract and technically challenging topic of ethical personal data management. The process resulted in participants reaching consensus on design considerations they deemed essential for realising a future with ethical personal data management. The Future workshop set the scene and encouraged a participatory atmosphere. The fantasy futures phase allowed us to creatively engage with the topic and opened up a space for new ideas. However, the Future workshop was limited in supporting an abstract problem space in which the expertise of participants and their alignment on the underlying concepts varied. Moreover, it was difficult to imagine radical alternatives to current technologies and data management practices.

Introducing Data Daemons turned out to be crucial. They immediately made the problem space tangible and participants took on the concept enthusiastically. While views on Data Daemons differed, the concept enabled participants to express their differences and align their perspectives on how they might interact with data ecosystems in mediated ways. Over time, perspectives converged and the Data Daemons became a tool for interrogating how end-users would like to relate to their personal tracking data. The group saw Data Daemons as a promising concept for considering ethical personal data management more comprehensively and with greater imagination. The Data Daemon concept itself merits further exploration. For example, we assumed that Data Daemons operate in the best interest of their owners. In real life this may of course not be the case and potential failures of Data Daemons should also be considered. 

Design fictions proved to be a useful tool for highlighting design considerations and expressing the Data Daemons concept. The narrative approach they offer made it possible for several participants to work collectively on a single design fiction, and to augment, merge, and refine the design fictions through iteration. Participants who did not actively create or refine a design fiction in a particular iteration were nonetheless able to critique it and offer additional insights and ideas based on the highlighted design considerations. The participatory process through which the design fictions and design considerations were developed gave legitimacy to the Data Daemon concept and the aspects of ethical personal data management that were elevated through it. We found this approach of using co-created narratives highly effective for gathering insights on how data processing and data protection guidelines and regulations may be put to practice ethically, in alignment with user values. 

In this study participants were self-selected, participated voluntarily, and had affiliations with research institutions or academia. In future work it will be important to assess the process's effectiveness for participants with limited knowledge of data or technology and citizens more broadly. A participatory process such as we present here also does not exist independently. Moving forward, we aim to embed it in established design practices, in particular in the conceptualisation phase of value-sensitive design where it has the potential to expand and contextualise conceptual and empirical investigations within the value-sensitive design process. 
\section{Conclusion}

In this study we have shown how a participatory approach involving design futures, a Future workshop, design fictions and the fictitious concept of Data Daemons, can empower non-expert end-users and designers to become active participants in the design process, capable of co-creating design considerations for ethical personal data management. The outcome is evident: it is not only possible, but necessary and valuable to include multiple perspectives and levels of expertise in considering this technically challenging and abstract topic. We hope this study inspires more research on participatory approaches to engaging end-users in the design of ethical personal data management.

\begin{acks}
This paper evolved from discussions at the First International Workshop on Ethical Data-Centric Design of Intelligent Behaviour at MobileHCI 2020. 
\end{acks}

\balance
\bibliographystyle{ACM-Reference-Format}
\bibliography{refs}

\end{document}